\documentclass[prd,twocolumn,showpcs,amsmath,amssymb,nofootinbib,preprintnumbers,balancelastpage]{revtex4}
\usepackage{epsfig}
\usepackage{amsmath}
\usepackage{bm}
\usepackage{times}
\usepackage{graphicx}
\usepackage{color}
\usepackage{slashed}
\usepackage{graphicx}
\usepackage{amsmath, amssymb}

\def\nn{\nonumber}
\def\bea{\begin{eqnarray}}
\def\eea{\end{eqnarray}}
\def\ba{\begin{eqnarray}}
\def\ea{\end{eqnarray}}
\def\be{\begin{equation}}
\def\ee{\end{equation}}

\begin{document}
\preprint{CALT 68-2858}

\title{Color octet scalars and high $p_T$ four-jet events at LHC}

\author{Jonathan M. Arnold and Bartosz Fornal\\
\textit{California Institute of Technology, Pasadena, CA 91125, USA}\\
}
\date{\today}

\begin{abstract}
We study the effect of color octet scalars on the high transverse momenta four-jet cross section at the LHC. We consider both
weak singlet and doublet scalars, concentrating on the case of small couplings to quarks. We find that a relatively early discovery at the LHC is possible for a range of scalar masses.
\end{abstract}

\maketitle
\bigskip

\section{Introduction}
If the scalar sector contains colored fields, there are two basic cases - either the scalars can couple to fermions or they cannot. If their gauge quantum numbers forbid couplings to fermions, then in order to enable the new scalars to decay they must be in a real representation of the color gauge group, which allows a cubic coupling in the scalar potential. The lowest dimensional real representation is a color octet. On the other hand, if the new colored scalars can couple to fermions, then one wants to impose minimal flavor violation (MFV) to forbid tree-level flavor changing neutral currents. If the new scalars are singlets under the flavor group, then the only representation that can couple to fermions is a color octet. So there are a number of reasons to focus on the color octet representation.

Color octet scalars appear in many models of new physics currently tested at the LHC. The literature on the subject is vast, covering the case of $\rm SU(2)$ singlets \cite{Bityukov:1997dh,Dobrescu3,Plehn:2008ae,Dobrescu,Choi:2009jc}, as well as doublets \cite{Manohar:2006ga, Gresham:2007ri,Gerbush:2007fe,Burgess:2009wm,Idilbi:2010rs,Kim:2008bx} and triplets \cite{Dobrescu2}. In this paper, we study the effect of color octet scalars on the high transverse momenta four-jet cross section at the LHC. We note that the analysis of multijet events is very promising \cite{Plehn:2008ae,Kilic:2008pm,Kilic:2008ub} and may give clues about new physics in the near future.

In the first part of the paper, we concentrate on color octet scalars with no weak quantum numbers. It is a simple extension of the standard model, naturally being anomaly-free and not affected by precision electroweak constraints.
The second part concerns weak doublet color octet scalars, namely the Manohar-Wise model \cite{Manohar:2006ga}.
Our interest in ${\rm SU}(2)$ doublets is motivated, in part, by the
observation that the principal of MFV
restricts gauge quantum numbers of any scalar sector coupled
directly to quarks \cite{Manohar:2006ga}.  The allowed quantum numbers are
those of the standard model Higgs doublet, or a color octet scalar
with the Higgs weak quantum numbers. In our analysis, we concentrate on cases with small couplings
to quarks.

For both weak singlets and doublets, we impose cuts on transverse momenta ($p_T$) of the jets, which
significantly reduces the standard model background. In the singlet case, we implement also
restrictions on the
invariant mass of jet pairs.
We find that, after performing such cuts, the color octet scalar contribution to
the four-jet cross section at the LHC is significant in a large region of parameter space.
Furthermore, we identify other weak doublet scalar signatures, which may be used to distinguish the doublets from
the singlets. However, for those additional processes involving weak doublets, the cross section depends also on other parameters in the scalar potential, which we keep fixed at certain values.

\section{Weak singlet signature}
We begin by investigating the standard model with
the addition of ${\rm SU}(2)$ singlet color octet scalars. Such scalars are
coupled to the standard model at tree-level only through the $\textrm{SU(3)}_{\rm c}$ gauge sector,
\begin{align}\label{1}
\hspace{-2mm} \mathcal{L}_{\rm kin}\!=\! \textrm{Tr}\left[\left(D_\mu S\right)^\dagger \!\left(D^\mu S\right)\right] \!=\! \frac{1}{2}(\partial_{\mu} S^a \!-\! g_s f^{a b c} G_{\mu}^b S^c)^2.
\end{align}
The self-coupling term enabling the scalar decay is,
\begin{align}\label{sss}
\mathcal{L}_{S S S} = -\frac{\mu\, M_s}{6} \,\textrm{Tr}\left(S^3\right),
\end{align}
where $M_s$ is the scalar mass.
The complete form of the scalar potential is given in ref.\,\,\cite{Bityukov:1997dh}.
Constraints considered there
yield $\mu \lesssim 1$.

\begin{figure}[t]
\centerline{\scalebox{0.47}{\includegraphics{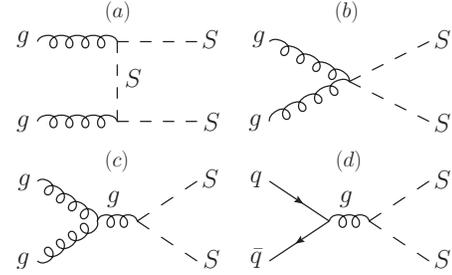}}}
\caption{\footnotesize{Feynman diagrams contributing to the $\,p \,p \rightarrow S \, S\,$ scattering.}}
\label{fig:FD1}
\end{figure}

\begin{figure}[t]
\centerline{\scalebox{0.40}{\includegraphics{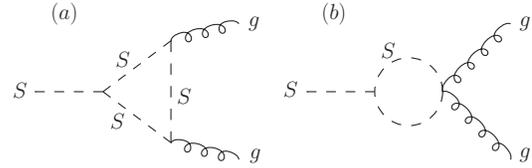}}}
\caption{\footnotesize{Diagrams representing color octet scalar decay $\,S \rightarrow g\,g$.
\\}}
\label{fig:FD2}
\end{figure}

We are interested in the effects of color octet scalars on the process $\,p \,p \rightarrow 4\,\rm jets$ at the LHC.
The cross section for such a process should be modified by the existence of the $\,p\, p \rightarrow S \, S \rightarrow g \,g \,g\, g\,$ channel. There are four tree-level diagrams contributing to  $\,p\, p \rightarrow S \, S\,$ scattering (figure \ref{fig:FD1}), while the scalar singlet decay, $\,S \rightarrow g\, g\,$, occurs only through loop diagrams (figure \ref{fig:FD2}).

The effective scalar-gluon-gluon vertex can be calculated from the diagrams in figure \ref{fig:FD2}. The corresponding effective Lagrangian term is,
\begin{align}\label{3}
\mathcal{L}^{\rm eff}_{S g g}=&\, \frac{\mu\,g_s^2}{128 \,\pi^2  M_s} \left(\frac{\pi^2}{9}-1 \right) \textrm{Tr}\left(G_{\mu \nu} G^{\mu \nu} S\right)\ ,
\end{align}
where $G_{\mu \nu}$ is the gluon field strength tensor. We note that decays of the weak singlet scalar to quarks (and gluons) can be induced by higher dimensional operators (for a detailed discussion, see refs. \cite{Dobrescu3,Dobrescu}). However, in our analysis we assume that the effect of such operators is negligible.

We expect the impact of $\rm SU(2)$ singlet color octet scalars on the standard model cross section for the process $\,p \, p \rightarrow 4\,\rm jets$ to be especially pronounced for high $p_T$ four-jet events, where the background is highly suppressed. In addition, appropriate cuts on the invariant mass of jet pairs should further increase the signal-to-background ratio. Our results are summarized in section IV A.

\section{Weak doublet signatures}
The second part of the paper is focused on signatures involving $\rm SU(2)$ doublet color octet scalars.  The motivation for such scalars was outlined in the Introduction. A thorough discussion is given in ref. \cite{Manohar:2006ga}.
Following this reference, we denote the new weak doublet scalar by,
\begin{align}
S^a = \left( \begin{array}{cc} {S^{+ a}} \\ {S^{0 a}} \end{array}\right),
\end{align}
where $a=1,\ldots,8$ is the index of the adjoint color representation.
One can split the neutral component of the new doublet into its real and imaginary parts,
\begin{align}
S^{0 a}={S^{0 a}_R+i S^{0 a}_I \over {\sqrt 2}}\ .
\end{align}
The tree-level masses are \cite{Manohar:2006ga},
\begin{align}
m^2_{S^{\pm}}&=m_S^2+\lambda_1{ v^2 \over 4} \ ,\nn \\
m^2_{S_R^0}&=m_S^2+\left(\lambda_1+\lambda_2+2\lambda_3\right){v^2 \over 4} \ ,\nn \\
m^2_{S_I^0}&=m_S^2+\left(\lambda_1+\lambda_2-2\lambda_3\right){v^2 \over 4}\ ,
\end{align}
where $m_S$ is the Lagrangian mass parameter, $\lambda_1, \lambda_2, \lambda_3$ are dimensionless parameters in the scalar potential, and $v$ is the vev of the standard model Higgs.
The full Lagrangian is given in ref.\,\,\cite{Manohar:2006ga}. Apart from the gauge coupling terms and the scalar potential,
one must also consider terms corresponding to the allowed couplings of color octet scalars to quarks.  In the quark mass eigenstate basis, the new Yukawa sector becomes,
\begin{align}\label{qL}
\mathcal{L}=&-\sqrt{2}\eta_U \bar u_R^i {m_U^i \over v}T^{a}u_L^i  S^{0 a}+\sqrt{2}\eta_U \bar u_R^i {m_U^i \over v}T^{a}V_{ij}d_L^j S^{+ a} \nn \\
&\!\!\!\!\!\!\!\!\!\!\!\!-\!\sqrt{2}\eta_D \bar d_R^i {m_D^i \over v}T^{a}d_L^i S^{0 \dagger a}\!-\!\sqrt{2}\eta_D \bar d_R^i {m_D^i \over v}V^{\dagger }_{ij}T^{a}u_L^j S^{- a}
\!\!+\!{\rm h.c.}
\end{align}
\begin{figure}[t]
\centerline{\scalebox{0.4}{\includegraphics{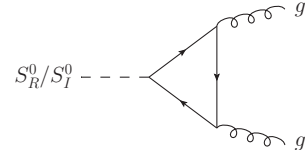}}}
\caption{\footnotesize{Diagram for a weak doublet color octet scalar decay to gluons through a quark loop.}}
\label{fig:FD3}
\end{figure}
\begin{figure}[t]
\centerline{\scalebox{0.55}{\includegraphics[trim = 0mm 4mm 0mm 0mm]{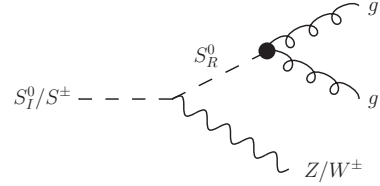}}}
\caption{\footnotesize{Tree-level diagram contributing to the weak doublet color octet scalar decays $\,S^0_I \rightarrow Z \,g\,g$ \ and $\,S^\pm \rightarrow W^\pm \,g\,g$.}}
\label{fig:FDn}
\end{figure}
\hspace{-1.0mm}Direct coupling to quarks is an important source of jet production in this model, but it is not the only one.  One must also consider decays of $S^0_R$ to gluon pairs through scalar loops (see, ref. \cite{Gresham:2007ri}) similar to those in figure \ref{fig:FD2}. This time, however, additional decays through quark loops (figure \ref{fig:FD3}) are allowed, with the diagram including a top loop contributing the most.

When color octet scalars are coupled to quarks through the Lagrangian terms (\ref{qL}) with large enough $\eta_U, \eta_D$, the dominant decay channel (for sufficiently large scalar masses) is to third-generation quarks, giving distinctive $\,t\, \bar{t}\, t\, \bar{t}\,$,
$\,t \,\bar{b}\, b \,\bar{t}\,$, and $\,b \,\bar{b}\, b\, \bar{b}\,$ signatures at the LHC (see, ref. \cite{Gerbush:2007fe}).  However, when the coupling to quarks is very small, the neutral real component $S^{0}_R$ decays predominantly to gluon pairs through scalar loops (the top quark loop contribution becomes negligible in this regime).
The effective Lagrangian term describing this decay is,
\begin{align}\label{3now}
\hspace{-1.7mm}\mathcal{L}^{\rm eff}_{S^0_R g g}=&\, \frac{3\,(\lambda_4+\lambda_5)\, v \,g_s^2}{64\,\pi^2  m_S^2} \left(\frac{\pi^2}{9}-1 \right) \textrm{Tr}\left(G_{\mu \nu} G^{\mu \nu} S^0_R\right),
\end{align}
where $\lambda_4$, $\lambda_5$ are two of the couplings in the scalar potential (see, eq. (6) in ref. \cite{Manohar:2006ga}).
In addition, the imaginary part of the neutral component $S^{0}_I$ and the charged components $S^{\pm}$ decay through diagrams shown in figure \ref{fig:FDn}.
We note that the decays $S^{0}_I \rightarrow Z\,g\,g\,$ and $S^{\pm} \rightarrow W^\pm\,g\,g\,$ can also occur through a scalar loop, but this effect turns out to be negligible.

In the case of small couplings to quarks, the weak doublet color octet scalars are produced similarly as the weak singlets (see, figure 1), with the scalar pairs being either: $S^0_R S^0_R$, $S^0_I S^0_I$, $S^+ S^-$, or $S^0_R S^0_I$. The scalar production in the first three cases has contributions from all diagrams shown in figure 1, whereas the last process is described only by the diagram in figure 1 (b).
Each of those cases corresponds to a different LHC signature: 4 jets, $2Z + 4\,\rm jets$, $W^+W^- + 4\,\rm jets$, and $Z+4\,\rm jets$, respectively.
Once again, the signal-to-background ratio can be improved by adopting high $p_T$ cuts.

\begin{figure*}[t]
\scalebox{0.95}{\includegraphics[trim = 6mm 0mm 0mm 0mm]{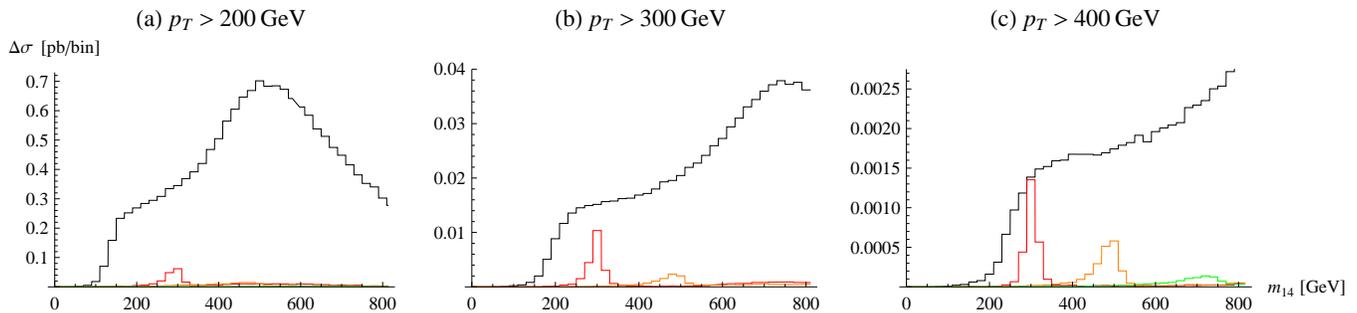}}
\caption{\footnotesize{The four-jet cross section per bin of the invariant mass of the highest and lowest $p_T$ jets for the invariant mass cut (9) and different $p_T$ cuts: (a) $200\ \rm GeV$, (b) $300\ \rm GeV$, and (c) $400\ \rm GeV$. The black curve is the standard model background. The colored curves: red, orange, and green, correspond to the signal of an $\rm SU(2)$ singlet color octet scalar of mass $300\ \rm GeV$, $500\ \rm GeV$, and $750\ \rm GeV$, respectively. The bin size is $20\ \rm GeV$.}}
\label{fig:scalarm14}
\end{figure*}

\section{Collider phenomenology}

We used MadGraph/MadEvent \cite{Alwall:2011uj} (version 1.3.1 of MadGraph 5) to simulate the standard model background and calculate the scalar signal cross section at the LHC running at 7 and 14 $\rm TeV$ center of mass energy, for different jet $p_T$ cuts and, in the weak singlet case, cuts on jet pair invariant masses. Apart from this, the default MadGraph run card was used, along with the $\rm cteq6L1$ PDFs. Jets were ordered by $p_T$, from highest to lowest. For the simulation of events involving scalars, we used FeynRules \cite{fuks,Christensen:2008py} to generate a model file for MadGraph/MadEvent. Events were run through Pythia and PGS.
MadAnalysis was used to plot the results.

We point out that a similar analysis for weak singlet scalars was performed in ref.\,\,\cite{Bityukov:1997dh} using Pythia 5.7. In this paper, however, we improve their analysis by running the events simulated using MadGraph/MadEvent through Pythia 6.420 and PGS4. We also implement a larger set of $p_T$ cuts and impose a slightly different invariant mass cut. In addition, we present plots of the differential cross section as a function of the invariant mass of jet pairs and discuss the current constraints on the color octet scalar masses.

We note that cuts can be tuned based on the scalar mass in order to maximize the effects of the scalars, at the same time
leaving a large enough data sample.
The restriction on the invariant mass of jet pairs we adopted is,
\begin{align}\label{masscut}
\min{ \left\{ \frac{\left| m_{12}-m_{34} \right|}{\left| \max{\{m_{12},m_{34}\}} \right|}, \mbox{\rm ~~perms.} \right\}} < 0.1\ ,
\end{align}
where $m_{i j}$ is the invariant mass of the $i$-th and $j$-th jet as ranked by $p_T$, and ``perms.'' denotes the other two possible pairings of the four jets.
This cut simply requires that for at least one of the possible pairings the resulting invariant masses of jet pairs are within $10\%$ of each other.
This should significantly suppress the standard model background, whereas the scalar signal is expected to essentially remain unchanged.

\subsection{Weak singlet}
Figure \ref{fig:scalarm14} shows the four-jet differential cross section as a function of the invariant mass of jets 1 and 4 for various weak singlet color octet scalar masses after adopting the jet invariant mass cut (\ref{masscut}) and different $p_T$ cuts. The signal is compared to the standard model background.
The cross sections corresponding to several invariant mass windows are given in table I.
In each case one can calculate the predicted number of events by multiplying the cross section by the integrated luminosity, and estimate the significance using the standard formula, $S = N_{\rm signal}/\sqrt{N_{\rm background}}$ .

\begin{figure}[t!]
\centerline{\scalebox{.86}{\includegraphics{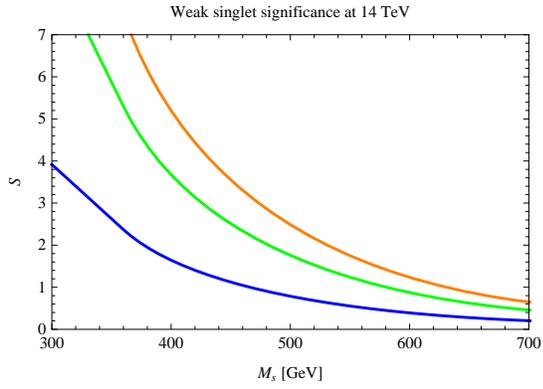}}}
\caption{\footnotesize{The significance of the $\rm SU(2)$ singlet color octet scalar four-jet signal within the invariant mass window $\,0.9 \ M_s\le m_{14} \le 1.1\  M_s\,$ as a function of the scalar mass for $E_{\rm CM}=14 \ \rm TeV$, $p_T > 200 \ \rm GeV$, and an integrated luminosity of $1 {\rm~ fb^{-1}}$ (blue), $5 {\rm~ fb^{-1}}$ (green), and $10 {\rm~ fb^{-1}}$ (orange).}}
\label{fig:muvspt}
\end{figure}

Figure \ref{fig:muvspt} shows the weak singlet scalar signal significance for different scalar masses and three sample data sizes at 14 TeV LHC center of mass energy. We choose $p_T > 200 \ \rm GeV$ since the discovery significance for higher $p_T$ cuts does not improve by much.
For scalar masses above 1 TeV the significance is too small for the signal to be detected with $10 \ \rm fb^{-1}$ of data. On the other hand, according to figure \ref{fig:muvspt}, if there exists a low mass color octet scalar, it should be discovered relatively early at the LHC running at 14 TeV.

\begin{figure}[t!]
\centerline{\scalebox{0.86}{\includegraphics{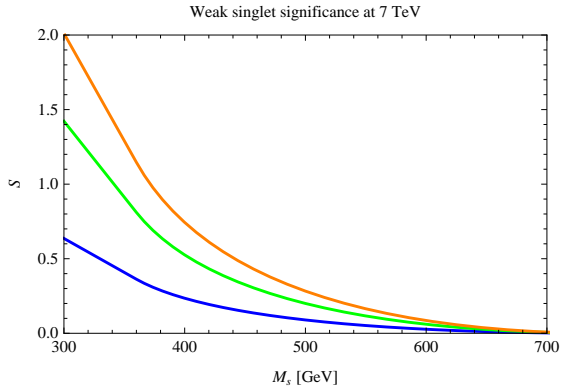}}}
\caption{\footnotesize{Same as figure \ref{fig:muvspt}, but for $E_{\rm CM}=7 \ \rm TeV$.
\\ }}
\label{fig:muvspt2}
\end{figure}

We also investigated current LHC constraints on the scalar masses with $1 \ \rm fb^{-1}$, $5 \ \rm fb^{-1}$, and $10 \ \rm fb^{-1}$ of data collected at $7\ \rm TeV$ center of mass energy.
Figure \ref{fig:muvspt2} shows the plot of the signal significance in this case as a function of the scalar mass.
Those data samples are still too small to extract any constraints on the color octet scalar masses from the four-jet analysis.

\begin{table}[h!]
\begin{center}
\begin{tabular}[t]{|c|c|c|c|c|c|}
  \hline
\multicolumn{2}{|c|}{} & \multicolumn{4}{|c|}{$M_s$} \\ \cline{3-6}
\multicolumn{2}{|c|}{\raisebox{2.0ex}[0pt]{$p_T^{\rm min}[\rm GeV]$}}  & \,\,\,$300\,{\rm GeV}$\,\, & \,\,\,$500\,{\rm GeV}$\,\,  & \,\,\,$750\,{\rm GeV}$\,\,  & \ \ \,\,$1\,{\rm TeV}$ \, \  \ \\
    \hline\hline
       \ \  &   $\sigma_{\scriptscriptstyle S} \, {\rm [fb]}$    & \, $1400$   \,  &\, $160$   \,&\, $16$   \,& \, $2.4$   \, \\
\ \raisebox{1.8ex}[0pt]{$100$}  \   &    $\,\sigma_{\scriptscriptstyle SM} \, {\rm [fb]}\,$    &  $1.9 \times 10^5$    &  $1.1 \times 10^5$    &  $5.3 \times 10^4$   &  $3.1 \times 10^4$    \\ \hline
       \ \  &   $\sigma_{\scriptscriptstyle S} \, {\rm [fb]}$    & \, $130$   \,&  \, $47$   \,& \, $10$   \,& \, $1.8$   \, \\
\ \raisebox{1.8ex}[0pt]{$200$}  \   &    $\sigma_{\scriptscriptstyle SM} \, {\rm [fb]}$    & \, $1100$   \, & \, $3500$   \, & \, $3000$   \, & \, $1700$   \, \\ \hline
       \ \  &   $\sigma_{\scriptscriptstyle S} \, {\rm [fb]}$    & \, $17$   \,  &\, $7.0$   \,&\, $3.4$   \,& \, $1.2$   \, \\
\ \raisebox{1.8ex}[0pt]{$300$}  \   &    $\sigma_{\scriptscriptstyle SM} \, {\rm [fb]}$    & \, $50$   \, & \, $100$   \, & \, $290$   \,& \, $260$   \, \\ \hline
       \ \  &   $\sigma_{\scriptscriptstyle S} \, {\rm [fb]}$    & \, $2.3$   \,&  \, $1.6$   \,& \, $0.66$   \,& \, $0.44$   \, \\
\ \raisebox{1.8ex}[0pt]{$400$}  \   &    $\sigma_{\scriptscriptstyle SM} \, {\rm [fb]}$    & \, $4.1$   \, & \, $8.7$   \, & \, $20$   \, & \, $42$   \, \\ \hline
\end{tabular}
\end{center}
\vspace{0mm}
\caption{\footnotesize{Cross sections for the weak singlet color octet scalar four-jet signal and the standard model background in the jet pair invariant mass window $\ 0.9 \ M_s\le m_{14} \le 1.1\  M_s\ $ for different $p_T$ cuts and scalar masses at $E_{\rm CM} = 14\ \rm TeV$.}}
\end{table}


\subsection{Weak doublet}

As was discussed in the previous section, the weak doublet color octet scalar signal depends on the strength of the couplings to quarks. Assuming $\,\lambda_4 + \lambda_5\simeq 1\,$, we find that for $\eta_U, \eta_D \gtrsim 10^{-6}$ the final states involving heavy quarks overwhelm the $\,4\,g\,$, $\,2 Z + 4\,g\,$, $\,W^+W^- + 4\,g\,$, and $\,Z + 4\,g\,$  final states. A detailed analysis of this case can be found in ref. \cite{Gerbush:2007fe}. On the other hand, when $\eta_U, \eta_D \lesssim 10^{-8}$, the four final states above become the dominant signatures. In our further analysis, we concentrate on the case of scalars decoupled from quarks, i.e., the limit $\eta_U, \eta_D \rightarrow 0$.

The four-jet signal resulting from the process $\,p\,p \rightarrow S^0_R\,S^0_R \rightarrow g\,g\,g\,g\,$ occurs through the same diagrams as in the weak singlet case (with $S^0_R$ instead of $S$). The branching ratio for $\,S^0_R\rightarrow g\,g\,$ is essentially one, therefore, all the results from the weak singlet section, including figures 5 -- 7, apply also here. After detecting such a signal, one would need to look for the other three signatures involving weak gauge bosons to distinguish the weak doublet case from the weak singlet color octet scalar signal.

\begin{figure}[t!]
\centerline{\scalebox{.92}{\includegraphics{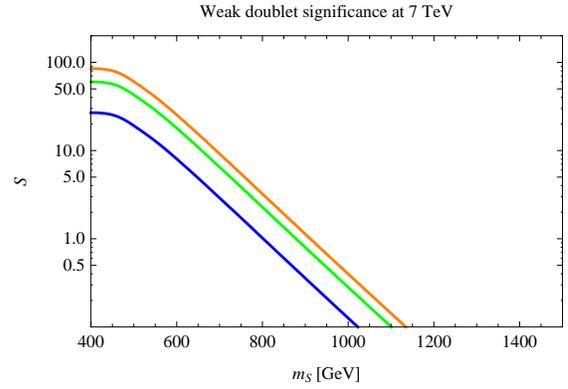}}}
\caption{\footnotesize{The logarithmic plot of the $\rm SU(2)$ doublet color octet scalar $2Z + 4\,\rm jets\,$ signal significance as a function of the scalar mass parameter $m_S$ for $\lambda_i = 1/2$, $E_{\rm CM}=7 \ \rm TeV$, $p_T > 100 \ \rm GeV$, and an integrated luminosity of $1 {\rm~ fb^{-1}}$ (blue), $5 {\rm~ fb^{-1}}$ (green), and $10 {\rm~ fb^{-1}}$ (orange).}}
\label{fig:muvspt33}
\end{figure}
\begin{figure}[t!]
\centerline{\scalebox{0.93}{\includegraphics{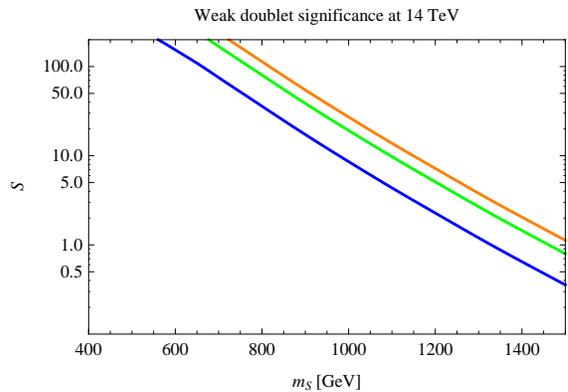}}}
\caption{\footnotesize{Same as figure \ref{fig:muvspt33}, but for $E_{\rm CM}=14 \ \rm TeV$.
\\ }}
\label{fig:muvspt233}
\end{figure}

The cross sections for the $2Z + 4\,\rm jets$, $W^+W^- + 4\,\rm jets$, and $Z+4\,\rm jets$ signals depend on the mass parameter $m_S$, as well as the values of $\lambda_1$, $\lambda_2$, and $\lambda_3$.
We emphasis that in our analysis of the $\rm SU(2)$ doublet case we choose particular values of $\lambda_i$, i.e., $\lambda_i = 1/2$, for $i = 1, 2, 3$, just to explain how the scalar search method we propose works. There are constraints on $\lambda_i$ coming from electroweak precision measurements and the requirement that the Higgs quartic coupling remain perturbative up to some high energy scale (see, ref.\,\,\cite{Gerbush:2007fe} for details). The values $\lambda_i = 1/2$ we adopted are consistent with those constraints. The only free parameter remaining is $m_S$. We note that for our choice of $\lambda_i$, the intermediate $S^0_R$ scalars in the decays $\,S^0_I \rightarrow Z\,S_R^0 \rightarrow Z \,g\,g\,$ and $\,S^\pm \rightarrow W^\pm \, S^0_R \rightarrow W^\pm \,g\,g\,$ are off-shell.

We find that the cross section for  $\,p\,p \rightarrow S^0_R\,S^0_I \rightarrow Z+4\,\rm jets\,$ is negligible compared to the other two processes.
Since we choose arbitrary values for $\lambda_i$, and because for a given choice of parameters
both processes $2Z + 4\,\rm jets\,$ and $W^+W^- + 4\,\rm jets\,$ have comparable cross sections, we limit our discussion  to the case $\,p\,p \rightarrow S^0_I\,S^0_I \rightarrow 2Z + 4\,\rm jets$. This channel is very promising since the standard model background is small.

Because of our arbitrary choice of $\lambda_i$ and the small background, it is natural to expect part of the $m_S$ parameter space to be already excluded by the currently collected data set from the LHC running at $E_{\rm CM} = 7 \ \rm TeV$. Figure \ref{fig:muvspt33} shows the plot of the $2Z + 4\,\rm jets\,$ signal significance as a function of $m_S$ for the integrated luminosities $1 \ \rm fb^{-1}$, $5 \ \rm fb^{-1}$, and $10 \ \rm fb^{-1}$ in this case.
Due to the challenge of simulating this standard model background, our analysis was restricted to the parton level.  A more detailed analysis would have to be done to set firm bounds, however, it is evident, at least in this approximation, that low mass weak doublet color octet scalars are already ruled out (for the particular choice of $\lambda_i = 1/2$) by many sigma. On the other hand, even $10 \ \rm fb^{-1}$ of data collected at 7 TeV center of mass energy still would not exclude the parameter region $m_S > 900 \ \rm GeV$.

Table II presents the $2Z + 4\,\rm jets\,$ scalar signal and the standard model background cross sections for different $p_T$ cuts and a few values of $m_S$ for $E_{\rm CM} = 14\ \rm TeV$. Imposing cuts on jet $p_T$ does not improve the discovery significance for $m_S < 800 \ \rm GeV$, but this parameter region is not interesting since it should already be excluded (for our choice of $\lambda_i$). For larger scalar masses the $p_T$ cuts can increase the significance. For example, in the case of $m_S \simeq 1.5 \ \rm TeV$ the significance increases by a factor of three when changing the $p_T$ cut from $100 \ \rm GeV$ to $200 \ \rm GeV$, and by a factor of five when changing the cut to $p_T > 300 \ \rm GeV$. Nevertheless, the signal itself drops then below five events for $10 \ \rm fb^{-1}$ of data.
Figure \ref{fig:muvspt233} shows the $2Z + 4\,\rm jets\,$ signal significance for 14 TeV center of mass energy as a function of $m_S$ for a few different integrated luminosities and $p_T > 100 \ \rm GeV$.
According to this plot, if there exists a weak doublet color octet scalar of mass $\sim 1 \ \rm TeV$, it should be discovered relatively early at the LHC running at $E_{\rm CM} = 14\ \rm TeV$. With $10 \ \rm fb^{-1}$ of data, a discovery of a scalar as heavy as $\sim 1.5\ \rm TeV$ might also be possible.
We note that cuts on the invariant mass of jet pairs would increase the significance of the signal, just as in the weak singlet case, but we leave this analysis for a future study.

\vspace{2mm}

\begin{table}[h!]
\begin{center}
\begin{tabular}[t]{|c|c|c|c|c|c|}
  \hline
    & \multicolumn{4}{|c|}{$\sigma_{m_S} \ [\rm fb]$} & \\ \cline{2-5}
    \raisebox{2.0ex}[0pt]{$p_T^{\rm min}[\rm GeV]$}  & \,$500\,{\rm GeV}$\, & \,$750\,{\rm GeV}$\,  & \ $1\,{\rm TeV}$ \,\  & \ $1.5\,{\rm TeV}$  \ & \raisebox{2.0ex}[0pt]{\ $\sigma_{\rm SM} \ [\rm fb]$ \ } \\ \cline{2-4}
    \hline\hline
\ \ $100$ \ \   &   \, $770$   \, & \, $150$   \, & \, $24$   \, & \, $1.0$   \,& \, $7.9$   \, \\
\ \ $200$ \ \   &   \, $18$   \, & \, $25$   \, & \, $11$   \, & \, $0.75$   \, & \, $0.48$   \, \\
\ \ $300$ \ \   &   \, $1.5$   \, & \, $1.6$   \, & \, $2.0$   \, & \, $0.43$   \, & \, $0.06$   \, \\ \hline
\end{tabular}
\end{center}
\vspace{-2mm}
\caption{\footnotesize{Cross sections for the weak doublet color octet scalar $2Z + 4\,\rm jets\,$ signal and the standard model background for different $p_T$ cuts and a few values of $m_S$ at $E_{\rm CM} = 14\ \rm TeV$.}}
\end{table}

\section{Conclusions}
We have investigated the effect of ${\rm SU(2)}$ singlet and doublet color octet scalars on high $p_T$ four-jet events at the LHC. We analyzed part of the available parameter space, concentrating on the region least explored so far, i.e., with no couplings or very weak couplings to quarks.
We identified the proper signatures in both cases and imposed cuts which improved the signal significance.

In case of weak singlet color octet scalars, one of the best signatures to look for are four-jet events. The standard model background is significantly suppressed by choosing high transverse momenta jets and cuts on the invariant mass of jet pairs. The signal is then strongly peaked around the invariant jet pair mass equal to the scalar mass. This method can be used to look for low mass scalars already in the first few inverse femtobarns of data from the LHC running at 14 TeV center of mass energy.

The same four-jet signature can be used to search for weak doublet color octet scalars with no couplings or very small couplings to quarks. However, in this case there are also other channels involving four jets accompanied by weak vector bosons in the final state. Because the standard model background is extremely small for those additional processes, such channels might be better to look at in search for weak doublet color octet scalars, especially since for massive scalars this signal should be much more significant than the four-jet signal. We performed such an analysis for particular values of the scalar potential parameters. In the case of stronger couplings between the $\rm SU(2)$ doublet scalars and quarks, final states involving bottom and top quarks are more promising channels for discovery.

\subsection*{Acknowledgment}
The authors would like to express their special thanks to Mark Wise for inspirational discussions and many extremely helpful comments at all stages of the work on this paper.
We are also very grateful to Johan Alwall for his continuous help, especially concerning the use of the MadGraph 5 software.
The work of the authors was supported in part by the U.S. Department of Energy under contract No. DE-FG02-92ER40701.
We further acknowledge the IISN ``MadGraph'' convention 4.4511.10 for access to the UCL cluster.
All Feynman
diagrams were drawn using JaxoDraw \cite{Binosi:2003yf}.



\begin{thebibliography}{99}


\bibitem{Bityukov:1997dh}
  S.~I.~Bityukov and N.~V.~Krasnikov,
  \textit{The search for new physics by the measurement of the four jet cross-section
  at LHC and FNAL},
  Mod.\ Phys.\ Lett.\  A {\bf 12}, 2011 (1997)
  [arXiv:hep-ph/9705338].

\bibitem{Dobrescu3}
  Y.~Bai, B.~A.~Dobrescu,
  \textit{Heavy octets and Tevatron signals with three or four b jets},
  JHEP {\bf 1107}, 100 (2011)
  [arXiv:1012.5814 [hep-ph]].


\bibitem{Plehn:2008ae}
  T.~Plehn and T.~M.~P.~Tait,
  \textit{Seeking sgluons},
  J.\ Phys.\ G {\bf 36}, 075001 (2009)
  [arXiv:0810.3919 [hep-ph]].


\bibitem{Dobrescu}
  B.~A.~Dobrescu, K.~Kong and R.~Mahbubani,
  \textit{Massive color-octet bosons and pairs of resonances at hadron colliders},
  Phys.\ Lett.\  B {\bf 670}, 119 (2008)
  [arXiv:0709.2378 [hep-ph]].

\bibitem{Choi:2009jc}
  S.~Y.~Choi, M.~Drees, J.~Kalinowski, J.~M.~Kim, E.~Popenda and P.~M.~Zerwas,
  \textit{Color-octet scalars at the LHC},
  Acta Phys.\ Polon.\  B {\bf 40}, 1947 (2009)
  [arXiv:0902.4706 [hep-ph]].


\bibitem{Manohar:2006ga}
 A.~V.~Manohar, M.~B.~Wise,
 \textit{Flavor changing neutral currents, an extended scalar sector, and
the Higgs production rate at the CERN Large Hadron Collider},
 Phys.\ Rev.\  {\bf D74}, 035009 (2006)
 [hep-ph/0606172].


\bibitem{Gresham:2007ri}
  M.~I.~Gresham and M.~B.~Wise,
  \textit{Color octet scalar production at the LHC},
  Phys.\ Rev.\  D {\bf 76}, 075003 (2007)
  [arXiv:0706.0909 [hep-ph]].

\bibitem{Gerbush:2007fe}
  M.~Gerbush, T.~J.~Khoo, D.~J.~Phalen, A.~Pierce, D.~Tucker-Smith,
  \textit{Color-octet scalars at the CERN LHC},
  Phys.\ Rev.\  {\bf D77}, 095003 (2008)
  [arXiv:0710.3133 [hep-ph]].






\bibitem{Burgess:2009wm}
  C.~P.~Burgess, M.~Trott and S.~Zuberi,
  \textit{Light octet scalars, a heavy Higgs and minimal flavour violation},
  JHEP {\bf 0909}, 082 (2009)
  [arXiv:0907.2696 [hep-ph]].



\bibitem{Kim:2008bx}
  C.~Kim and T.~Mehen,
  \textit{Color octet scalar bound states at the LHC},
  Phys.\ Rev.\  D {\bf 79}, 035011 (2009)
  [arXiv:0812.0307 [hep-ph]].




\bibitem{Idilbi:2010rs}
  A.~Idilbi, C.~Kim and T.~Mehen,
  \textit{Pair production of color octet scalars at the LHC},
  Phys.\ Rev.\  D {\bf 82}, 075017 (2010)
  [arXiv:1007.0865 [hep-ph]].

\vspace{0.7mm}

\bibitem{Dobrescu2}
  B.~A.~Dobrescu and G.~Z.~Krnjaic,
  \textit{Weak-triplet, color-octet scalars and the CDF dijet excess},
  [arXiv:1104.2893 [hep-ph]].


\bibitem{Kilic:2008pm}
  C.~Kilic, T.~Okui and R.~Sundrum,
  \textit{Colored resonances at the Tevatron: Phenomenology and discovery potential
  in multijets},
  JHEP {\bf 0807}, 038 (2008)
  [arXiv:0802.2568 [hep-ph]].

\bibitem{Kilic:2008ub}
  C.~Kilic, S.~Schumann and M.~Son,
  \textit{Searching for multijet resonances at the LHC},
  JHEP {\bf 0904}, 128 (2009)
  [arXiv:0810.5542 [hep-ph]].




\bibitem{Alwall:2011uj}
  J.~Alwall, M.~Herquet, F.~Maltoni, O.~Mattelaer and T.~Stelzer,
  \textit{MadGraph 5: Going beyond},
  JHEP {\bf 1106}, 128 (2011)
  [arXiv:1106.0522 [hep-ph]].

\bibitem{Christensen:2008py}
  N.~D.~Christensen and C.~Duhr,
  \textit{FeynRules - Feynman rules made easy},
  Comput.\ Phys.\ Commun.\  {\bf 180}, 1614 (2009)
  [arXiv:0806.4194 [hep-ph]].

\bibitem{fuks}
  C.~Degrande, C.~Duhr, B.~Fuks, D.~Grellscheid, O.~Mattelaer and T.~Reiter,
  \textit{UFO - The Universal FeynRules Output},
  arXiv:1108.2040 [hep-ph].


\bibitem{Binosi:2003yf}
  D.~Binosi and L.~Theussl,
  \textit{JaxoDraw: A graphical user interface for drawing Feynman diagrams},
  Comput.\ Phys.\ Commun.\  {\bf 161}, 76 (2004)
  [hep-ph/0309015].










\end{thebibliography}
\end{document}